\documentclass[12pt]{article}
\usepackage{verbatim,color,amssymb,bbm,epsfig,amsmath}
\usepackage[usenames,dvipsnames]{xcolor}
\usepackage{fancyhdr}
\usepackage[authoryear, sort]{natbib}
\usepackage{wasysym, bbm}
\usepackage{url}

\newtheorem{theorem}{\underline{\bf Theorem}}
\newtheorem{assumption}{\underline{\bf Assumption}}

\newtheorem{lemma}{\underline{\bf Lemma}}

\def\D{{\cal D}}

\def\boxit#1{\vbox{\hrule\hbox{\vrule\kern6pt  \vbox{\kern6pt#1\kern6pt}\kern6pt\vrule}\hrule}}

\def\rcom#1 {{\color{red}\bf#1} }

\def\sumi{\hbox{$\sum_{i=1}^n$}}

\def\diag{\hbox{diag}}

\def\diag{\hbox{diag}}
\def\log{\hbox{log}}

\def\bse{\begin{eqnarray*}}
\def\ese{\end{eqnarray*}}
\def\be{\begin{eqnarray}}
\def\ee{\end{eqnarray}}
\def\bq{\begin{equation}}
\def\eq{\end{equation}}
\def\bse{\begin{eqnarray*}}
\def\ese{\end{eqnarray*}}
\def\pr{\hbox{pr}}

\def\trans{^{\rm T}}

\def\th{^{th}}
\def\b1e{{\mathbf e}}

\def\trans{^{\rm T}}

\def\th{^{th}}
\def\b1e{{\mathbf e}}

\renewcommand{\hat}{\widehat}
\renewcommand{\tilde}{\widetilde}

\newcommand{\nn}{n^{-1}}
\newcommand{\hth}{\hat{\theta}}

\newcommand{\hb}{\hat{b}}
\newcommand{\hbe}{\hat{\beta}}
\newcommand{\tbe}{\tilde{\beta}}
\newcommand{\hw}{\hat{W}}
\def\md{ {\cal D}}

\newcommand{\tth}{\widetilde \theta_{km\D'}}
\newcommand{\p}{\mathrm{pr}}
\def\si{\hbox{$\sum_{i=1}^n$}}
\def\sm{\hbox{$\sum_{m=1}^M$}}
\def\sk{\hbox{$\sum_{k=1}^K$}}
\def\th{^{\rm th}}
\def\mm{\mathcal{M}}
\def\ttho{\widetilde \theta_{k_0m_0\D'}}
\def\bth{\bar{\theta}_{\md'}}
\def\hv{\hat{V}_{km\md}}

\makeatletter

\def\trans{^{\rm T}}
\def\t{\rm T}
\def\boxit#1{\vbox{\hrule\hbox{\vrule\kern6pt  \vbox{\kern6pt#1\kern6pt}\kern6pt\vrule}\hrule}}

\def\bse{\begin{eqnarray*}}
\def\ese{\end{eqnarray*}}
\def\be{\begin{eqnarray}}
\def\ee{\end{eqnarray}}

\begin{document}

\thispagestyle{empty}
\baselineskip=28pt

\begin{center}
{\LARGE{\bf Data integration in high dimension \\ with multiple quantiles}}
\end{center}

\baselineskip=12pt

\vskip 2mm
\begin{center}
Guorong Dai and Ursula U. M\"uller\\
Department of Statistics, Texas A\&M University, College Station, TX 77843, USA \\
rondai@stat.tamu.edu and uschi@stat.tamu.edu\\
\hskip 5mm \\
\hskip 5mm\\ \hskip 5mm\\
Raymond J. Carroll\\
Department of Statistics, Texas A\&M University, College Station, TX 77843, USA \\ and School of Mathematical and Physical Sciences, University of Technology Sydney, Broadway NSW 2007, Australia\\
carroll@stat.tamu.edu
\end{center}

\begin{center}
{\Large{\bf Abstract}}
\end{center}

This article deals with the analysis of high dimensional data that come from multiple sources (``experiments") and thus have different possibly correlated responses, but share the same set of predictors. The measurements of the predictors may be different across experiments. We introduce a new regression approach with multiple quantiles to select those predictors that affect any of the responses at any quantile level and estimate the nonzero parameters. Our estimator is a minimizer of a penalized objective function, which aggregates the data from the different experiments. We establish model selection consistency and asymptotic normality of the estimator. In addition we present an information criterion, which can also be used for consistent model selection. Simulations and two data applications illustrate the advantages of our method, which takes the group structure induced by the predictors across experiments and quantile levels into account.

\baselineskip=12pt
\par\vfill\noindent
\underline{\bf Some Key Words}:  Data integration; High dimensional data; Information criterion; Penalized quantile regression.

\par\medskip\noindent
\underline{\bf Short title}: Data integration with multiple quantiles

\clearpage\pagebreak\newpage
\pagenumbering{arabic}
\newlength{\gnat}
\setlength{\gnat}{22pt}
\baselineskip=\gnat

\section{Introduction}\label{sec1}

	To set the stage for this work on data integration, first consider $K$  different data sets with linear regression models
\be
Y_{k} = X_{k}\trans\alpha^*_k+U_{k} \quad (k=1,\ldots,K).
\label{model0}
\ee
Here $Y_{k}$ is a scalar response, $X_{k}$ is a $p$-dimensional predictor, $\alpha^*_k$ is a $p$-dimensional parameter vector and $U_k$ is the error term. \citet{zellner1962efficient} referred to this set of models as {\em seemingly unrelated regressions} and proposed the idea of estimating the regression parameters simultaneously using a generalized least squares method.
The responses in model (\ref{model0}) are different, but dependent, while the predictors are the same in the $K$ data sets, but not their values. This is, for example, given if individuals are assessed through various responses from different experiments and the predictor values are measured in different ways \citep{g&c2017}.

Model (\ref{model0}), with the assumption that $E(U_k\mid X_k)=0$, can also be written as a heterogenous linear regression model, i.e., as
\bse
E(Y_k - X_k\trans \alpha_k^*\mid X_k )  =0 \quad( k=1,\ldots,K).
\ese
We consider the same scenario, but pursue a different approach. Instead of modeling the conditional mean of the response given the covariates, we assume linear regression models for the conditional quantiles $Q_{\tau_m}(X_k)$ at various quantile levels $\tau_m$ ($m=1,\ldots,M$), i.e., 
\be \label{modelQR}
E\{ I(Y_k \le X_k\trans\theta_{km}^*) - \tau_m \mid X_k \} = 0 \quad (k=1,\ldots,K),
\ee
where $I(\cdot)$ is the indicator function and $\theta^*_{km}$ is a $p$-dimensional parameter vector. This is equivalent to
\bse
\p(Y_k \le X_k\trans\theta_{km}^* \mid X_k) =\p\{Y_k \le Q_{\tau_m}(X_k) \mid X_k\}
=  \tau_m\ (m=1,\ldots,M;\  k=1,\ldots,K).
\ese

We are interested in the high dimensional data situation and therefore let the dimension $p=p_n$ of the parameter vector tend to infinity as the sample size $n$ increases.
In addition, we assume that the data are sparse, i.e.\ most of the parameters are zero, which means that only a fraction of the predictors affect the responses.

An important goal is to identify the relevant predictors. One possible approach is to aggregate each predictor's effect in all experiments by forming groups. In our scenario all responses share the same set of predictors. Hence we have a natural group structure: the parameters of different quantiles and experiments that belong to the same predictor constitute a group; see \citet{g&c2017}, who developed a group penalized estimation method using a pseudolikelihood. To handle the unspecified dependence between the responses in the $K$ experiments, they pooled the marginal likelihoods and imposed $L_2$-group penalization on the grouped parameters. The group penalty was introduced in a 1999 Australian National University Ph.D. thesis by S. Bakin and then applied to group selection questions by \citet{yuan2006model}. \citet{g&c2017} used it to select predictors that are influential in any of the experiments. The main tool in their article is the smoothly clipped absolute deviation penalty \citep{fan2001variable}. In addition, \citet{g&c2017} used the concept of the Bayesian information criterion to also develop a pseudolikelihood information criterion that applies to the high dimensional scenario. The pseudolikelihood approach they employed is an important advance and useful when the distribution of the error can be modeled parametrically, which is not assumed in our case.

In this article we use a linear quantile regression approach based on model (\ref{modelQR}),
i.e.\ we will not work with a likelihood, but with a different objective function.
Quantile regression was introduced by \citet{koenker1978regression}; see also \citet{k2005}. In contrast to classical regression, it provides a global picture of the predictors' effect on the distribution of the responses, while it is robust to heavy-tailed distributions. In high dimensional settings \citet{belloni2011} studied linear quantile regression with a Lasso penalty, \citet{wang2012quantile} proved selection consistency of linear quantile regression with nonconvex penalty functions, and \citet{sherwood2016partially} derived asymptotic properties of partially linear additive quantile regression with a nonconvex penalty. In addition to these articles on single quantile regression, \citet{zou2008composite} introduced a composite quantile regression approach for linear models, which considers multiple quantiles simultaneously. They assumed that the slopes were the same across quantiles and used the adaptive Lasso penalty from \citet{zou2006adaptive}. The method shares the oracle properties proposed in \citet{fan2001variable}. In the presence of heterogeneity, i.e.\ when the covariates and the error are dependent so that the slopes vary across quantiles, the method of \citet{zou2008regularized} is able to detect non-zero slopes simultaneously. \citet{zou2008regularized}  generalized the approach to the case with multiple responses. The two 2008 articles by Zou \& Yuan consider only the scenario with a fixed number of parameters. Moreover, \citet{fan2016multitask} studied quantile regression with multiple responses under the assumption that the responses and predictors can be transformed to a multivariate normal variable by some monotone function, which is not posited in our model. Unlike us who are interested in identifying relevant predictors, they focused on predicting responses and estimating correlation matrices.


Our goal is simultaneous variable selection with multiple quantiles  across $K$ experiments. To take account of the unknown dependence structure between the responses in the different experiments, we integrate the data by summing up their quantile loss functions. Additionally, similar to \citet{sherwood2016partially} who conducted variable selection with multiple quantiles, we apply a nonconvex penalty on the $L_1$-norm of the coefficients related to each predictor, which represents the overall strength of the predictor across multiple experiments and quantiles. This penalty function takes the group structure into account and excludes covariates that have no impact on any of the responses at any of the quantile levels. Moreover, the $L_1$-norm is computationally convenient in quantile regression settings, thanks to \citet{peng2015iterative}, who provided a new ``Quick Iterative Coordinate Descent'' algorithm for solving nonconvex penalized quantile regression in high dimensions with no group structure. With modifications, their algorithm can be adapted to our approach; see Section \nolinebreak \ref{sec4}.

Multiple quantile regression for dependent data that originate from different sources has, to the best of our knowledge, not been studied in the literature. Apart from this we also cover the high dimensional data scenario by adding a nonconvex group penalty term. We establish selection consistency and asymptotic normality of our estimator in this quite general setting under mild assumptions. Additionally we propose a multiple quantile Bayesian information criterion (MQBIC) based on pooled check functions, which is an extension of the Bayesian information criterion for linear quantile regression \citep{lee2014model} to the multiple experiment scenario. Similar to the pseudolikelihood information criterion in \citet{g&c2017}, MQBIC permits consistent model selection (see Section \ref{sec3}) and choice of the tuning parameter for the penalized estimator (see Section \ref{sec4}).

Summing up, the main contribution of this article is the introduction of quantile based methods to the high dimensional scenario of data integration. We propose a penalized estimation process and an information criterion, which can identify the covariates that affect any of the responses at any of the quantile levels. Our method enjoys robustness and can be applied to the complex scenario with heterogeneous data and dependent responses.

The rest of this article is organized as follows. In Section \ref{sec2} we introduce our objective function, which involves a nonconvex group penalization term, and present the oracle properties of the estimator. The MQBIC is presented in Section \ref{sec3} and its model selection consistency is established. In Section \ref{sec4} we compare our method with other approaches using simulations. Our method is illustrated in Section \ref{sec5} by means of empirical data examples. Section \ref{sec6} gives a brief conclusion of the article and a discussion of further questions. All proofs are in the Appendix. For notational clarity we assume in the following that the sample sizes and the quantile levels are the same in every experiment. The conclusions and methods are essentially the same if we drop these assumptions.

\section{Penalized estimator}\label{sec2}
Throughout this article we will use the capital letter $C$ to represent a generic constant, including $C_1$, $C_2$, etc. We write $I_m$ for the $m\times m$ identity matrix. The symbols $\|\cdot\|_1$ and $\|\cdot\|$ refer to the $L_1$- and $L_2$- norms of a vector and $\otimes$ denotes the Kronecker product.

Our conditional quantile regression model is $Q_{\tau_m}(X_{k})=X_{k}\trans\theta_{km}^*$ with ordered levels $0<\tau_1<\tau_2<\dots<\tau_M<1$. We can set the first column of $X_k$ to be $(1,\ldots,1)\trans$ so that the model contains intercept terms. For notational convenience, we assume the intercepts all equal zero. The number of predictors $p_n$ tends to infinity as the sample size $n$ increases.

For $k=1,\ldots,K$ and $i=1,\dots,n$ we consider $n$ independent copies $\{Y_{ki},X_{ki}\}$ with $X_{ki}=(X_{ki1},\dots,X_{kip_n})\trans$ of the base observation $\{Y_k,X_k\}$  from model (\ref{model0}). Here we use three subscripts to locate the predictors, i.e.\ $X_{kij}$ represents the $j\th$ component of the $i\th$ observation in the $k\th$ experiment. We write $X_{k\cdot j}=(X_{k1j},\dots,X_{knj})\trans$ for the vector. The data are summarized in Table \ref{table0}.

The regression parameters $\theta^*_{km}$ ($k=1,\ldots,K$, $m=1,\ldots,M$) are assumed to be sparse, i.e.\ most of the components of $\theta^*_{km}$ are zero. Write $\theta^{*(j)}$ for the parameters related to the $j\th$ predictor ($j =1,\ldots, p_n$) across the $K$ experiments and the $M$ quantile levels, i.e.\  
$\theta^{*(j)}=(\theta^*_{11j},\dots,\theta^*_{1Mj},\dots,\theta^*_{K1j},\dots,\theta^*_{KMj})\trans$. 
We want to select the predictors that have an effect on any of the responses, i.e.\ we want to specify the set $\mathcal{A}=\{j:1\leq j\leq p_n,\|\theta^{*(j)}\|>0\}$. Without loss of generality let $\mathcal{A}=\{1,2,\dots,q_n\}$, i.e.\ only the first $q_n$ predictors have nonzero parameters. We assume that $q_n$ tends to infinity as $n$ and $p_n$ increase. For convenience of notation, we use the letter $a$ at the end of a subscript if we refer to subvectors or submatrices that consist of components with subscripts in $\mathcal{A}$. For example, $X_{kia}=(X_{ki1},\dots,X_{kiq_n})\trans$, $X_{k\cdot a}=(X_{k1a},\dots,X_{kna})\trans$ and $\theta^*_{kma}=(\theta^*_{km1},\dots,\theta^*_{kmq_n})\trans$.

\begin{table}
	\renewcommand\arraystretch{1.5}
	\centering
	\caption{\baselineskip=12pt Data structure of multiple experiments}
	\label{table0}
	\begin{tabular}{|cccc|}
		\hline
		& Experiment 1& \ldots& Experiment $K$ \\*[-.30em]
		\hline
		Parameters of $\tau_1$ & $\theta_{11}^*=(\theta_{111}^*,\ldots,\theta_{11p_n}^*)\trans$& \ldots& $\theta_{K1}^*=(\theta_{K11}^*,\ldots,\theta_{K1p_n}^*)\trans$ \\*[-.60em]
		\vdots& \vdots& & \vdots \\*[-.60em]
		Parameters of $\tau_M$ & $\theta_{1M}^*=(\theta_{1M1}^*,\ldots,\theta_{1Mp_n}^*)\trans$& \ldots& $\theta_{KM}^*=(\theta_{KM1}^*,\ldots,\theta_{KMp_n}^*)\trans$ \\*[-.60em]
		Observation 1 & $Y_{11}$, $X_{11}=(X_{111},\ldots,X_{11p_n})\trans$ & \ldots& $Y_{K1}$,  $X_{K1}=(X_{K11},\ldots,X_{K1p_n})\trans$ \\*[-.60em]
		\vdots& \vdots& & \vdots \\*[-.60em]
		Observation $n$& $Y_{1n}$, $X_{1n}=(X_{1n1},\ldots,X_{1np_n})\trans$ & \ldots& $Y_{Kn}$,  $X_{Kn}=(X_{Kn1},\ldots,X_{Knp_n})\trans$ \\
			\hline
	\end{tabular}
	\begin{quote}
		\footnotesize
	\end{quote}
\end{table}

The dependence between the experiments is unspecified. To integrate the data we therefore sum up the quantile loss functions across the $K$ experiments and the $M$ quantiles, 
\be
\ell_n(\theta)=\nn\sk\sm\si\rho_m(Y_{ki}-X_{ki}\trans\theta_{km}).
\label{loss}
\ee
Here $\rho_m(x)=x\{\tau_m-I(x<0)\}$ is the check function and $\theta=(\theta_{11}\trans, \dots, \theta_{1M}\trans,\dots,\theta_{K1}\trans,\dots,\theta_{KM}\trans)\trans$ is a parameter vector. To select the predictors that affect any of the responses, a nonconvex penalty function $\Omega_{\lambda_n}(\cdot)$ with tuning parameter $\lambda_n$ is imposed on the overall impact of each predictor.
That impact is represented by the $L_1$ norm of the vector $\theta^{(j)}$,
which contains the parameters of the $j\th$ predictor in the $K$ experiments.
This gives the overall objective function
\be
\Gamma_{\lambda_n}(\theta)=\ell_n(\theta)+\hbox{$\sum^{p_n}_{j=1}$}\Omega_{\lambda_n}(\|\theta^{(j)}\|_1). \label{obhetero}
\ee
Our estimator is obtained by minimizing $\Gamma_{\lambda_n}(\theta)$. We use the smoothly clipped absolute deviation (SCAD) penalty function \citep{fan2001variable}
$$ 
\Omega_{\lambda _n}(x)
=
\lambda_nxI(0\leq x\leq\lambda_n)+\frac{a\lambda_nx-(x^2+\lambda_n^2)/2}{a-1}I(\lambda_n< x<a\lambda_n)+\frac{(a+1)\lambda_n^2}{2}I(x\geq a\lambda_n),
$$ 
where $a$ is a constant that is usually set to 3.7 \citep{fan2001variable}. Before stating the asymptotic properties of our estimator, we make the following assumptions. 
\begin{assumption}\label{AA1}
There is a constant $C>0$ such that $|X_{kij}|\leq C$ for every $k =1,\ldots, K$, $i=1,\ldots, n$ and $j = 1,\ldots, p_n$.
\end{assumption}

\begin{assumption}\label{AA2}
	For every $k=1,\ldots,K$ there are positive constants $C_1$ and $C_2$ such that
	\bse
	C_1\leq\lambda_{\min}(n^{-1}X\trans_{k\cdot a}X_{k\cdot a})\leq\lambda_{\max}(n^{-1}X\trans_{k\cdot a}X_{k\cdot a})\leq C_2,
	\ese
	where $\lambda_{\min}(\cdot)$ and $\lambda_{\max}(\cdot)$ stand for the smallest and the largest eigenvalue, respectively. In addition, the true model contains at least one continuous covariate, and $X_{k\cdot a}$ and $(Y_{k1},\ldots,Y_{kn})\trans$ are in ``general positions", which is an identifiability condition that guarantees that a solution to the quantile regression problem exists \citep[Section 2.2.2]{k2005}.
\end{assumption}

\begin{assumption}\label{AA3}
	For every $k=1,\ldots,K$ and $m=1,\ldots,M$, the conditional probability density $f_{km}(\cdot\mid x)$ of $\varepsilon_{km}=Y_{k}-X_{k}\trans\theta_{km}^*$ given $X_k=x$ is uniformly bounded and bounded away from zero in a neighborhood of zero, and has a derivative $f_{km}'(\cdot\mid x)$, which is uniformly bounded in a neighborhood of zero.
\end{assumption}


\begin{assumption}\label{AA4}
  The true model size satisfies $q_n=O(n^{c_1})$ for some $0\leq c_1<1/2$.
\end{assumption}

\begin{assumption}\label{AA5}
	 There are positive constants $c_2$ and $C$ such that $2c_1<c_2\leq 1$, where $c_1$ is the constant introduced in Assumption \ref{AA4}, and $n^{(1-c_2)/2}\min_{1\leq j\leq q_n}\|\theta^{*(j)}\|_1\geq C.$
\end{assumption}
Assumptions \ref{AA1} and \ref{AA2} guarantee good behavior of the design matrices.
The conditions in Assumption \ref{AA3} concern the unknown distribution of the random errors. They are considerably weaker than assuming a specific parametric model for the error distribution. Assumption \ref{AA4} regulates the growth rate of the true model size. This is a standard assumption for linear models with a diverging number of parameters; see, for example, \citet{wang2012quantile} and \citet{lee2014model}. Assumption \ref{AA5} excludes situations where the nonzero parameters decay too fast. Conditions similar to Assumptions \ref{AA1}--\ref{AA5} were required in \citet{wang2012quantile} for single experiments with a single quantile.

The oracle estimator $\hat{\theta}$ is defined as the minimizer of $\ell_n(\theta)$ that knows that the first $q_n$ components of $\theta$ are nonzero and that the others are zero, i.e.\  $\|\hat{\theta}^{(j)}\|=0$ for $q_n<j\leq p_n$. The following theorem provides the  model selection consistency of our estimator. More precisely, we will show that, with probability tending to one, the oracle estimator can be obtained with our approach, i.e.\ by minimizing the objective function $\Gamma_{\lambda_n}(\theta)$.

\begin{theorem} \label{theorem 1}
Let $S(\lambda_n)$ denote the set of local minimizers of $\Gamma_{\lambda_n}(\theta)$ and $\hat \theta$ the oracle estimator. Under Assumptions \ref{AA1}--\ref{AA5}, $\p\{\hth\in S(\lambda_n)\}\to1$ as $n\to \infty$, if $\lambda_n=o\{n^{-(1-c_2)/2}\}$, $n^{-1/2}q_n=o(\lambda_n)$ and $n^{-1}\log\, p_n=o(\lambda_n^2)$.
\end{theorem}

The next theorem, Theorem \ref{theorem 2}, gives the asymptotic normality of the nonzero part of the oracle estimator $\hth$ from Theorem \ref{theorem 1}, i.e.\ of $\hth_a$. We first introduce some notation. For $k=1,\ldots, K$, $m=1\,\ldots,M$ and $i=1,\ldots,n$ we write
\bse
&&\varepsilon_{kmi}=Y_{ki}-X_{ki}\trans\theta^*_{km},\  \varepsilon_{km}=(\varepsilon_{km1},\dots,\varepsilon_{kmn})\trans,\ \varepsilon=(\varepsilon_{11}\trans,\dots,\varepsilon_{1M}\trans,\dots,\varepsilon_{K1}\trans,\dots,\varepsilon_{KM}\trans)\trans, \nonumber \\
&&\psi_{kmi}(\varepsilon)=\tau_m-I(\varepsilon_{kmi}<0),\ \psi_{nkm}(\varepsilon)=\{\psi_{km1}(\varepsilon),\ldots,\psi_{kmn}(\varepsilon)\}\trans, \label{psi} \\
&&\psi_{nk}(\varepsilon)=\{\psi_{nk1}(\varepsilon)\trans,\ldots,\psi_{nkM}(\varepsilon)\trans\}\trans, \ \psi_n(\varepsilon)=\{\psi_{n1}(\varepsilon)\trans,\ldots,\psi_{nK}(\varepsilon)\trans\}\trans, \nonumber \\  &&H_n=E\{\psi_n(\varepsilon)\psi_n(\varepsilon)\trans\mid \mathcal{X}\}\ \hbox{with $\mathcal{X}=\{X_{ki}:k=1,\dots,K,i=1,\dots,n\}$}, \nonumber \\
&&B_{nkm}=\diag\{f_{km}(0\mid X_{k1}),\dots,f_{km}(0\mid X_{kn})\},\ B_{nk}=\diag(B_{nk1},\dots,B_{nkM}), \label{bnk} \\
&&B_n=\diag(B_{n1},\dots,B_{nK}), \ \theta_a^*=(\theta_{11a}^{*\t},\dots,\theta_{1Ma}^{*\t},\dots,\theta_{K1a}^{*\t},\dots,\theta_{KMa}^{*\t})\trans,\\ 
&&\hth_{kma}=(\hth_{km1},\dots,\hth_{kmq_n})\trans, \ \hth_a=(\hth_{11a}\trans,\dots,\hth_{1Ma}\trans,\dots,\hth_{K1a}\trans,\dots,\hth_{KMa}\trans)\trans. \nonumber
\ese

\begin{theorem} \label{theorem 2}
Let $n^*=n\times M\times K$, $q^*_n=q_n\times M\times K$. Denote $X_a=\diag(I_M\otimes X_{1\cdot a},\dots, I_M\otimes X_{K\cdot a})$ as a $n^*\times q_n^*$ block diagonal matrix, $R_n=n^{-1}X_a\trans B_nX_a$, $S_n=n^{-1}X_a\trans H_nX_a$ and $\Sigma_n=R_n^{-1}S_nR_n^{-1}$. Consider a $s\times q_n^*$ matrix $A_n$ with $s$ fixed and $A_n A_n\trans\to G$, a positive definite matrix, then
\bse
n^{1/2}A_n\Sigma_n^{-1/2}(\hat{\theta}_a-\theta_a^*)\to N(0,G) \quad (n \to \infty)
\ese
in distribution, provided Assumptions \ref{AA1}-\ref{AA4} are satisfied and $\lambda_{\min}(S_n)$ is uniformly bounded away from zero.
\end{theorem}

Theorems \ref{theorem 1} and \ref{theorem 2} establish the model selection consistency and asymptotic normality of our estimator when experiments are correlated. This shows that it is reasonable to aggregate information from multiple experiments, rather than ignoring the correlation and analyzing each experiment separately.

\section{Multiple quantile Bayesian information criterion}\label{sec3}

To select the correct model we use an information criterion that balances the goodness-of-fit and the complexity of a model. By applying this information criterion to a set of competing models, the true model can be identified with probability approaching one. In the context of quantile regression, \citet{lee2014model} developed a Bayesian information criterion with a diverging number of predictors. That method considers one single quantile and deals with data from one single experiment.
We use a generalized version of the criterion, now based on multiple quantiles and on data from several experiments, which improves its ability to select the correct model.

The {\em multiple quantile} Bayesian information criterion of a submodel $\md\subset\{1,2,\dots,p_n\}$ is
\be
\hbox{MQBIC}(\md)=\log\{\hbox{$\sk\sm\si$}\rho_m(Y_{ki}-X_{ki\md}\trans\hth_{km\md})\}+(2n)^{-1}|\md|T_n\log\,n,
\label{mq}
\ee
where $\hat{\theta}_{km\md}=\mathop{\arg\min}_{\theta\in\mathbb{R}^{|\md|}}\sum_{i=1}^n\rho_m(Y_{ki}-X_{ki\md}\trans\theta)$ for $k=1,\dots,K$ and $m=1,\dots,M$, $|\md|$ is the cardinality of $\md$, and $T_n$ is a sequence of positive constants diverging to infinity as $n$ increases. The notation $X_{ki\md}$ refers to the subvectors of $X_{ki\cdot}$ which only contain the components with subscripts in $\md$. We set an upper bound on the cardinality of competing models, say $d_n$, and search for the best model among submodels whose cardinality is smaller or equal to $d_n$. Define $\md^*=\{1, 2, \dots, q_n\}$ as the subset of $\{1,\dots, p_n\}$ corresponding to the true model, and $\mathcal{M}=\{\md\subset\{1, \dots, p_n\}:|\md|\leq d_n\}$ as the set of all competing models. The first part of the MQBIC represents the goodness-of-fit, while the second term is a penalty on the model complexity. To guarantee  model selection consistency of the MQBIC we need the following assumptions, in addition to some of the assumptions from Section \ref{sec2}.
	
\begin{assumption}\label{AB2}
	For every $k=1,\ldots,K$ there are constants $0<C_3\leq C_4$ such that for any $\md\subset\{1, \dots, p_n\}$ the matrix $X_{k\cdot\md}=(X_{k1\md},\dots,X_{kn\md})\trans$ satisfies
	\bse
	C_3\leq\hbox{$\min_{|\md|\leq 2d_n}$}\lambda_{\min}(n^{-1}X\trans_{k\cdot \md}X_{k\cdot \md})\leq\hbox{$\max_{|\md|\leq 2d_n}$}\lambda_{\max}(n^{-1}X\trans_{k\cdot \md}X_{k\cdot \md})\leq C_4.
	\ese
\end{assumption}
	
\begin{assumption}\label{AB4}
	The full model size $p_n$ is of order $p_n=O(n^{c_3})$ for some $c_3>0$; the true model size $q_n$ is fixed, $q_n=q$, and satisfies $q\leq d_n=O(n^{c_4})$ for some $0<c_4<1/2$.
\end{assumption}

\begin{assumption}\label{AB5}	
	The sequence $T_n$ in the definition (\ref{mq}) satisfies $T_n\to\infty$ and $n^{-1}T_n\log\,n\to0$.
\end{assumption}

\begin{assumption}\label{AB6}
	The average of the check functions, $n^{-1}\sum^K_{k=1}\sum_{m=1}^M\sum_{i=1}^n\rho_m(\varepsilon_{kmi})$, is bounded and bounded away from zero with probability tending to one.
\end{assumption}	
Assumption \ref{AB2} extends Assumption \ref{AA2} for the true model to all candidate models. This is common for scenarios with more regression parameters than observations, i.e.\ $p_n>n$. In Assumption \ref{AB4}, the true model size is fixed because of a technical difficulty in handling the maximum of $|\md\backslash\md^*|^{-1}|\nn\si\{\rho_m(Y_{ki}-X_{ki\md}\trans\hth_{km\md})-\rho_m(Y_{ki}-X_{ki\md^*}\trans\hth_{km\md^*})\}|$ over the set of overfitted models $\{\md\in\mm:\md^*\subset\md$, $\md\neq\md^*\}$ \citep{lee2014model}. Assumption \ref{AB5} regulates the growth rate of the sequence $T_n$. Assumption \ref{AB6} is made for convenience in the proofs because $n^{-1}\sum^K_{k=1}\sum_{m=1}^M\sum_{i=1}^n\rho_m(\varepsilon_{kmi})$ appears in denominators.

In the following theorem we show that the true model has, with probability tending to one, the smallest MQBIC value among all candidate models. 

\begin{theorem}\label{theorem 3}
	 If Assumptions \ref{AA1}, \ref{AA3} and \ref{AB2}-\ref{AB6} hold, then with probability tending to one, the true model can be selected by minimizing the MQBIC, that is
	 \bse
	 \hbox{$\lim_{n\to\infty}$}\p\{\hbox{$\min_{\md\in(\mm\backslash \{\md^*\})}$}\hbox{MQBIC}(\md)>\hbox{MQBIC}(\md^*)\}=1.
	 \ese
\end{theorem}
Theorem \ref{theorem 3} establishes model selection consistency of the MQBIC for data from multiple dependent sources, which provides another approach to identify the true underlying model.  In the MQBIC approach estimation and model selection are separate processes. This is different from minimizing the objective function in Section \ref{sec2}, which is a one-step procedure. The main advantage of the MQBIC is that we can use it to select the tuning parameter $\lambda_n$ for the penalized estimation process in Section \ref{sec2}, which is computationally more efficient than cross validation. The details are given in Section \ref{sec4}.

\section{Simulations}\label{sec4}
In this section we study the numerical performance of our estimators. We use the objective function (\ref{obhetero}) with $M=5$ quantiles, $\tau_1=1/6,\tau_2=2/6,\ldots,\tau_5=5/6$, and study two different group structures, namely complete and incomplete grouping. Complete grouping means that parameters of the same predictor can only be either all zero or all nonzero, while in the incomplete case a group may contain both zero and nonzero predictors.

In both cases the number of experiments is $K=2$, the sample size is $n=100$ and the number of predictors is $p=100$ or $p=200$. The nonzero parameters are drawn independently from a uniform distribution on $[0.05,1]$. For $K=1,2$ we generate independent random vectors $X'_{ki}$, $i=1,\ldots,100$, from a $p$-dimensional multivariate normal distribution with mean zero and a covariance matrix whose $(i,j)\th$ component is $0.5^{|i-j|}$ for $1\leq i,j\leq p$. The predictors $X_{ki}$ for the different scenarios described below are transformations of the $X'_{ki}$'s. For $i=1,\dots,100$ the error terms $(\xi_{1i},\xi_{2i})\trans$ are drawn independently from a bivariate normal distribution with mean zero or from a bivariate t distribution with three degrees of freedom. The covariance matrix of  $(\xi_{1i},\xi_{2i})$  is $\Sigma$ with entries $\Sigma_{11}=\Sigma_{22}=1$ and $\Sigma_{12}=\Sigma_{21}=0.7$. For minimizing the objective functions we use an algorithm by \citet{peng2015iterative}, modified for multiple quantiles and experiments. The majorization function in that article \citep[equation (7)]{peng2015iterative} becomes $\nn\sk\sm\si\rho_m(Y_{ki}-X\trans_{ki}\theta_{km})+\sum_{j=1}^{p_n}\Omega_{\lambda_n}'(\|\tilde{\theta}^{(j)}\|_1+)\|\theta^{(j)}\|_1$. Here $\Omega'_{\lambda_n}(\cdot)$ is the derivative of $\Omega_{\lambda_n}(\cdot)$; $\tilde{\theta}$ is the result from the previous iteration. The minimization of the modified majorization function can be done using the algorithm in Section 3 of \citet{peng2015iterative}. We refer to that article for a detailed description. The tuning parameter $\lambda$ is chosen from a grid $\Lambda$. For $\lambda\in\Lambda$ let $\hat{\theta}_{\lambda, km}=(\hat{\theta}_{\lambda, km1},\dots,\hat{\theta}_{\lambda, kmp})\trans$ denote the estimators obtained from minimizing the objective function (\ref{obhetero}) with $\lambda_n=\lambda$, where $k=1,2$ and $m=1,2,3,4,5$. Further let $\md_\lambda=\{j:1\leq j \leq p,\sum^K_{k=1}\sum^M_{m=1}|\hat{\theta}_{\lambda, kmj}|>0\}$. In order to obtain the final estimator we use
\be
\hat{\lambda}=\hbox{$\mathop{\arg\min}_{\lambda\in\Lambda}$}\big[\log\, \{\sk\sm\si\rho_m(Y_{ki}-X_{ki}\trans\hat{\theta}_{\lambda, km}) \}+(2n)^{-1}|\md_\lambda|(\log\,n)T\big],
\label{mqbic}
\ee
which minimizes the MQBIC. This approach adapts criterion (2.10) in \citet{lee2014model} to multiple quantile levels and experiments. Since that article recommends $T=C\, \log\,p$ and their simulation results show this type of information criterions tends to underfit models slightly, we consider $T=(\log\,p)/3$ or $(\log\,p)/6$ and examine how this affects the performance of the method. In each scenario we record the following three indices.
\begin{enumerate}
	\item Positive selection rate (PSR): the proportion of selected predictors that affect any quantile of any response. Then, formally, PSR $=$  $|\hat{\mathcal{A}}\cap\mathcal{A}|/|\mathcal{A}|$ with $\mathcal{A}=\{j:1\leq j\leq p, \|\theta^{*(j)}\|>0\}$ and $\hat{\mathcal{A}}=\{j:1\leq j\leq p, \|\hth^{(j)}\|>0\}$.
	
	\item False discovery rate (FDR): the proportion of selected predictors that affect no response, i.e.\ $|\hat{\mathcal{A}}\cap\mathcal{A}^c|/|\mathcal{A}^c|$.
	
	\item Absolute error (AE): the absolute estimation error, i.e.\ $(KM)^{-1}\|\hth-\theta^*\|_1$.
\end{enumerate}

\begin{table}
	\centering
	\caption{\baselineskip=12pt Positive selection rates, false discovery rates and absolute errors of the data integration method and the combined analysis for models with normal errors and complete group structure. Here DI denotes the data integration method, CA-$\tau$ the combined analysis with one quantile $\tau=$2/6 or 3/6; PSR is the positive selection rate, FDR the false discovery rate and AE the absolute error $(KM)^{-1}\|\hth-\theta^*\|_1$. The parameter $T$ in criterion (\ref{mqbic}) is (a) $(\log\,p)/3$ or (b) $(\log\,p)/6$.
	}
	\vskip 3mm
	\label{table1}
	\begin{tabular}{|llrrrrrrr|}
		\hline
		& &  \multicolumn{3}{c}{$p=100$} & & \multicolumn{3}{c|}{$p=200$}  \\*[-.30em]
		& & PSR(\%) & FDR(\%) & AE & & PSR(\%) & FDR(\%) & AE  \\
		& DI & 98.3 (5.0) & 1.1 (1.5) & 0.3 (0.1) &  & 98.2 (5.2) & 0.7 (0.7)  & 0.3 (0.1)  \\*[-.60em]
		(a) & CA-(2/6) & 83.3 (7.5) & 2.4 (2.2) & 0.6 (0.1) &  & 78.0 (8.2) & 0.8 (0.7) & 0.7 (0.1)  \\*[-.60em]
		& CA-(3/6) & 81.7 (5.0) & 1.4 (1.4) & 0.3 (0.1) &  & 79.2 (7.3) & 0.7 (0.7) & 0.3 (0.1)  \\*[-.60em]
		&&&&&&&&\\*[-.60em]
		& DI & 99.0 (4.0) & 1.9 (2.4) & 0.2 (0.1) &  & 98.3 (4.0) & 1.1 (0.1)  & 0.3 (0.1)  \\*[-.60em]
		(b) & CA-(2/6) & 92.3 (8.7) & 19.2 (16.2) & 0.8 (0.3) &  & 89.3 (5.6) & 28.1 (0.7) & 1.5 (0.7)  \\*[-.60em]
		& CA-(3/6) & 83.3 (4.1) & 6.9 (8.7) & 0.3 (0.2) &  & 88.7 (1.6) & 12.2 (0.3) & 0.6 (0.5)  \\
		\hline
	\end{tabular}
\end{table}

\begin{table}
	\centering
	\caption{\baselineskip=12pt We consider the same scenario as Table \ref{table1}, but now the predictors have an incomplete group structure.}
	\vskip 3mm
	\label{table2}
	\begin{tabular}{|llrrrrrrr|}
		\hline
		& &  \multicolumn{3}{c}{$p=100$} & & \multicolumn{3}{c|}{$p=200$}  \\*[-.30em]
		& & PSR(\%) & FDR(\%) & AE & & PSR(\%) & FDR(\%) & AE  \\
		& DI                 & 97.2 (5.6) & 1.8 (1.7) & 0.4 (0.1) &  & 91.3 (9.7) & 0.8 (0.9)  & 0.4 (0.1)  \\*[-.60em]
		(a) & CA-(2/6)      & 86.0 (6.8) & 3.4 (3.0) & 0.7 (0.1) &  & 82.9 (6.0) & 1.4 (1.2) & 0.8 (0.1)  \\*[-.60em]
		& CA-(3/6) & 84.6 (5.4) & 2.2 (1.9) & 0.4 (0.1) &  & 83.8 (6.2) & 1.1 (1.0) & 0.4 (0.1)  \\*[-.60em]
		&&&&&&&&\\*[-.60em]
		& DI & 98.0 (4.3) & 2.4 (2.1) & 0.3 (0.2) &  & 96.6 (6.5) & 2.0 (2.0)  & 0.4 (0.1)  \\*[-.60em]
		(b) & CA-(2/6) & 92.2 (7.3) & 23.7 (16.5) & 0.9 (0.3) &  & 92.0 (7.1) & 32.6 (18.3) & 1.7 (0.7)  \\*[-.60em]
		& CA-(3/6) & 87.2 (4.8) & 7.6 (8.6) & 0.4 (0.2) &  & 87.1 (7.3) & 13.7 (16.5) & 0.8 (0.6)  \\
		\hline
	\end{tabular}
\end{table}

The data integration (DI) approach is compared with the standard method, a combined analysis based on the $\tau\th$ quantile (CA-$\tau$). That method considers only one quantile, $\tau$. It analyzes the data from the two experiments separately and then merges the two sets of selected predictors. We will see that in most of the cases the CA-$\tau$ method selects more unimportant predictors than the DI approach. This indicates that the false discovery rate will rise even further when the results from different quantile levels are combined. We therefore did not consider this approach. In Tables \ref{table1}-\ref{table3} we present the average values of the three indices calculated from $100$ simulated data sets. The standard deviations are provided in parentheses. 

Table \ref{table1} shows the simulation results for a scenario with normal errors and complete group structure. The nonzero parameters are $\alpha_{11}^*$, $\alpha_{16}^*$, $\alpha_{1(12)}^*$, $\alpha_{1(15)}^*$, $\alpha_{1(20)}^*$ and $\alpha_{21}^*$, $\alpha_{26}^*$, $\alpha_{2(12)}^*$, $\alpha_{2(15)}^*$, $\alpha_{2(20)}^*$. Let $\Phi(\cdot)$ be the distribution function of a standard normal variable. For $k=1,2$ and $i=1,\dots,100$ the predictors are $X_{ki3}=\Phi(X_{ki3}')$ and $X_{kij}=X_{kij}'$ for $j\neq 3$. The responses are $Y_{ki}=X_{ki}\trans\alpha^*_k+0.7\xi_{ki}X_{ki3}$. The DI method achieves the highest positive selection rates and the lowest false discovery rates. It also has the lowest absolute errors. Apparently the DI method is not much affected by the choice of $T$.

\begin{table}
	\centering
	\caption{\baselineskip=12pt We consider the scenario from Table \ref{table2} with an imcomplete group structure, but now the random errors follow a bivariate t distribution with three degrees of freedom.}
	\vskip 3mm
	\label{table3}
	\begin{tabular}{|llrrrrrrr|}
		\hline
		& &  \multicolumn{3}{c}{$p=100$} & & \multicolumn{3}{c|}{$p=200$}  \\*[-.30em]
		& & PSR(\%) & FDR(\%) & AE & & PSR(\%) & FDR(\%) & AE  \\
		& DI                 & 93.7 (6.9) & 1.4 (1.4) & 0.5 (0.1) &  & 89.7 (9.9) & 0.7 (0.8)  & 0.5 (0.2)  \\*[-.60em]
		(a) & CA-(2/6)      & 83.0 (6.8) & 2.6 (2.6) & 0.8 (0.1) &  & 80.7 (6.6) & 1.4 (1.5) & 0.9 (0.2)  \\*[-.60em]
		& CA-(3/6) & 81.2 (5.8) & 1.7 (1.8) & 0.5 (0.1) &  & 81.3 (6.5) & 0.9 (0.8) & 0.5 (0.1)  \\*[-.60em]
		&&&&&&&&\\*[-.60em]
		& DI & 94.9 (6.0) & 2.0 (2.2) & 0.4 (0.1) &  & 94.1 (7.3) & 1.7 (1.7)  & 0.5 (0.1)  \\*[-.60em]
		(b) & CA-(2/6) & 88.7 (8.0) & 12.9 (13.0) & 0.8 (0.3) &  & 85.0 (8.4) & 12.7 (15.7) & 1.3 (0.8)  \\*[-.60em]
		& CA-(3/6) & 84.8 (5.6) & 5.4 (5.6) & 0.4 (0.2) &  & 83.7 (6.4) & 4.8 (9.7) & 0.6 (0.5)  \\
		\hline
	\end{tabular}
\end{table}

In Tables \ref{table2} and \ref{table3} we present the simulation results for the same scenario as in the previous table, but now the predictors have an {\em incomplete} group structure. The error variables in the two tables have a normal distribution (Table \ref{table2}) and a t distribution with three degrees of freedom (Table \ref{table3}).  The nonzero parameters are $\alpha_{14}^*$, $\alpha_{16}^*$, $\alpha_{19}^*$, $\alpha_{1(12)}^*$, $\alpha_{1(15)}^*$, $\alpha_{1(20)}^*$ and $\alpha_{21}^*$, $\alpha_{26}^*$, $\alpha_{2(12)}^*$, $\alpha_{2(15)}^*$, $\alpha_{2(20)}^*$, $\alpha_{2(25)}^*$. For $i=1,\dots,100$ the predictors in the first experiment are $X_{1i1}=\Phi(X_{1i1}')$ and $X_{1ij}=X_{1ij}'$ for $j\neq 1$. The predictors in the second experiment are $X_{2i3}=\Phi(X_{2i3}')$ and $X_{2ij}=X_{2ij}'$ for $j\neq 3$. The responses are $Y_{1i}=X_{1i}\trans\alpha^*_1+0.7\xi_{1i}X_{1i1}$ and $Y_{2i}=X_{2i}\trans\alpha^*_2+0.7\xi_{2i}X_{2i3}$. Inspecting the quantities in the two tables we see that the DI again has higher positive selection rates and lower false discovery rates. Also it produces similar or smaller absolute errors than its competitors. We observe that in both tables criterion (\ref{mqbic}) using $T=(\log\, p)/6$ selects larger models compared with that using $T=(\log\, p)/3$. The results in Table \ref{table3} also illustrate the robustness of quantile regression when dealing with heavy-tailed distributions. For the t error distribution we omit the results for the simpler case with completely grouped predictors, where our approach also works well.

\section{Examples}\label{sec5}
\subsection{Multiple experiments}\label{sec5.1}

In this section we apply our method to data from a liver toxicity study \citep{bushel2007simultaneous}, which are avaliable in the {\tt R} package {\tt mixOmics} \citep{mixomics}. In the study two groups of 32 male rats each were exposed to non-toxic (50 or 150 mg/kg) and toxic (1,500 or 2,000 mg/kg) doses of acetaminophen (paracetamol), respectively. There is a data set for each group, which contains the rats' expression profiles of 3,116 genes and level of cholesterol. Due to the different experimental environments, the two data sets have different measurements. We want to identify the genes that significantly affect the response, namely the level of cholesterol on a logarithmic scale, based on aggregating the two data sets. To preprocess the data the genes are sorted by the absolute values of their correlation coefficients with the response in each set. The top 50 genes in each set are retained as covariates in the analysis.

To fit sparse models, we minimize the objective function (\ref{obhetero}) using all data. We consider quantiles  $\tau_m=m/10$ for $m=1,\ldots,9$ and use two different penalties, the SCAD penalty and the minimax concave penalty (MCP). The tuning parameters of the penalties are chosen using formula (\ref{mqbic}), i.e.\ as minimizers of the MQBIC, with $T=\log\, p/6$. In addition, we take an approach based on random partitions: we divide each data set randomly into two parts, a training set of size 24 and a validation set of size 8. This is repeated 50 times. The training set is used to select parameters and obtain parameter estimates as before, i.e.,\ by   minimizing (\ref{obhetero}) with $\lambda$ chosen using (\ref{mqbic}). The prediction errors $\sk\sm\si\rho_m(Y_{ki}-X_{ki}\trans\hat{\theta}_{km}-\hat{b}_{km})$ are calculated based on the estimates from the training sets and data $X,Y$ from the validation sets. Here $\hb_{km}$ is the estimated intercept in the conditional quantile $Q_{\tau_m}(X_k)$. For comparison we also consider the combined analysis, which treats the data sets separately and then combines the results. We record the sizes of the models that are fitted using the entire data sets, and the simulated means and standard deviations of the model sizes and prediction errors otained from the 50 replications. 

\begin{table}
	\centering
	\caption{\baselineskip=12pt Analysis of the liver toxicity data. The sizes of the selected subset models (column 2) are based on all data, the average sizes and prediction errors (column 3 and 4) are based on the data using random partitions. The standard deviations are in parentheses. Here DI denotes the data integration method, CA the combined analysis, SCAD the smoothly clipped absolute deviation and MCP the minimax concave penalty.}
	\vskip 5mm
	\label{tableliver}
	\begin{tabular}{|lrrr|}
		\hline
		& All Data & \multicolumn{2}{c|}{Random Partition} \\*[-.60em]
		& Model Size &  Model Size & {Prediction error} \\*[-.30em]
		DI with SCAD & 4 & 3.12 (1.61) & 1.82 (0.72)   \\*[-.60em]
		DI with MCP & 3 & 3.04 (1.54) &  1.85 (1.00)  \\*[-.60em]
		CA with SCAD & 6 & 6.72 (2.56) &  1.97 (0.72)  \\*[-.60em]
		CA with MCP & 10 & 7.64 (3.37) &  1.98 (0.79)  \\
		\hline
	\end{tabular}
\end{table}

Table \ref{tableliver} shows the results of analyzing the liver toxicity data. When using the entire data sets, the DI method with SCAD penalty selects 4 covariates, which include the 3 covariates selected by the DI method with MCP and are also chosen by the combined analysis with either of the two penalties. Using the random partition approach, the DI method generates models that are, on average, more sparse than those obtained from the combined analysis, with lower prediction errors.

\subsection{Multiple responses}\label{sec5.2}

As a second application, now with a multivariate response vector, we analyze data sets of financial market indices from the {\tt R} package {\tt FusionLearn} \citep{fusionlearn}. These data contain three correlated indices: the VIX index, the S\&P 500 index and the Dow Jones index. The VIX and the S\&P 500 are negatively correlated, while the S\&P 500 and the Dow Jones are positively correlated \citep{g&c2017}. The covariates are 46 major international equity indices, North American bond indices and major commodity indices. In the analysis the transformation $\log(V_t$\ /\,$V_y)\times 100$ of each index is used, where $V_t$ and $V_y$ denote today's and yesterday's value. The training data set consists of 232 records of three years' market performances with three-day spacing between the values. As shown in \citet{g&c2017}, the values are not autocorrelated at a 5\% significance level.

\begin{table}
	\centering
	\caption{\baselineskip=12pt Analysis of the financial market indices. The figures are the prediction errors and the sizes of the selected submodels. The full model size is $p=46$. Here DI denotes the data integration method, CA the combined analysis, UR is unpenalized regression, SCAD denotes smoothly clipped absolute deviation and MCP minimax concave penalty.}
	\vskip 5mm
	\label{table5}
	\begin{tabular}{|lrrrr|}
		\hline
		& Model Size & \multicolumn{3}{c|}{Prediction errors}   \\*[-.60em]
		& & VIX & S\&P 500 & Dow Jones     \\*[-.30em]
		DI with SCAD &  4 & 10045.8 & 524.9 & 306.9  \\*[-.60em]
		DI with MCP &  4 & 10026.5 & 522.7 & 308.8   \\*[-.60em]
		CA with SCAD &  23 & 10139.9 & 637.6 & 398.6  \\*[-.60em]
		CA with MCP &  19 & 10115.8 & 637.8 & 391.0  \\*[-.60em]
		UR &  46 & 13408.5 & 644.0 & 663.4  \\
		\hline
	\end{tabular}
	\begin{quote}
		\footnotesize
	\end{quote}
\end{table}

As before, we minimize the objective function (\ref{obhetero}) to select covariates and estimate parameters. The quantiles in (\ref{obhetero}) are $\tau_m=m/20$ for $m=1, 2, \dots, 19$. We again use the SCAD penalty and the  MCP, and determine their tuning parameters with criterion (\ref{mqbic}). The SCAD penalty selects 4 covariates, which are the same as the 4 covariates selected by the MCP penalty. The competing methods are the combined analysis with the two penalties and unpenalized regression. The latter includes all 46 covariates in the model and generates estimators by minimizing the loss function (\ref{loss}) without a penalty term. We use the five fitted models for predictions based on a (different) validation data set with 464 records. Prediction errors for the three indices, that is $\sum_{m=1}^M\sum^n_{i=1}\rho_m(Y_{ki}-X_{ki}\trans\hat{\theta}_{km}-\hat{b}_{km})$ for $k=1,2,3$, are recorded in Table \ref{table5}. The DI method with both the SCAD penalty and the MCP outperforms the other three approaches, while DI with the SCAD penalty and DI with the MCP yields similar prediction errors. Apart from that, the DI method yields models that are considerably smaller than those from the combined analysis, i.e.\ it achieves more sparsity. The two empirical data examples in Sections \ref{sec5.1} and \ref{sec5.2} again clearly demonstrate the advantages of our method.

\section{Conclusion and discussion}\label{sec6}

To the best of our knowledge we are the first to introduce a quantile regression approach to a data integration scenario with high dimensional data. By considering multiple quantiles simultaneously we obtain a global picture of the relationship between predictors and responses. A penalized estimator and an information criterion, which aggregate information from multiple experiments, were developed to select variables and to estimate model parameters. Our method copes with heterogeneity in the data. It successfully exploits the group structure in the parameter set across quantiles and experiments so that influential predictors can be identified.

In practice quality and relevance of data may vary from one source to another. Therefore a weighted version of the loss function (\ref{loss}),
\bse
\ell_n^{(w)}(\theta)=\nn\sk w_k\sm\si\rho_m(Y_{ki}-X_{ki}\trans\theta_{km}),
\ese
with weight vector $w=(w_1,\dots,w_K)\trans$, may improve our estimator, which uses uniform weights. It would be worthwhile to specify and construct such weights for data from different experiments.

The nonconvex penalty function associated with the $L_1$-norm has different properties compared to the penalty function associated with the $L_2$-norm employed by \citet{g&c2017}, which forces parameters in the same group to be all zero or all nonzero. When the least squares approach is used, \citet{jiang2014concave} show that the penalty associated with the $L_1$-norm can be applied if the group structure is incomplete, i.e.,\ both zero and nonzero parameters exist in the same group, which is called ``bi-level selection'' property. In this article we focus on groups of parameters to identify predictors that have an impact on one or more responses at some quantile levels. In the simulations of Section \ref{sec4} we saw that the SCAD penalty with the $L_1$-norm actually performs well at the group level even if the group structure is incomplete.  Theoretical properties of the $L_1$-norm in the quantile regression setting, however, still need to be investigated in greater detail.

\baselineskip=16pt
\section*{Supplementary material}
\begin{itemize}
	\item All the programs  of Section \ref{sec4} and \ref{sec5} are available at
	\url{https://github.com/guorongdai/Data-Integration}. 
	
	\item The data in Section \ref{sec5.1} are from the {\tt R} package {\tt FusionLearn}, while the data in Section \ref{sec5.2} are from the {\tt R} package {\tt mixOmics}.
\end{itemize}

\section*{Acknowledgments}
Dai and Carroll's research was supported by a grant from the National Cancer Institute (U01-CA057030).

\setcounter{equation}{0}
\setcounter{section}{0}
\renewcommand{\thesection}{\Alph{section}}
\renewcommand{\thesubsection}{\thesection.\arabic{subsection}}
\renewcommand{\theequation}{\thesection.\arabic{equation}}

\section{Appendix}\label{seca}

\begin{lemma}\label{lemma 0}
Use the notation from Section \ref{sec2} and write
\bse
\tbe_{nkm}=n^{1/2}(X_{k\cdot a}\trans B_{nkm} X_{k\cdot a})^{-1}X\trans_{k\cdot a}\psi_{nkm}(\varepsilon)
\ese
 for $k=1,\dots,K$ and $m=1,\dots,M$. Then, provided Assumptions \ref{AA1}, \ref{AA2}, \ref{AA3} and \ref{AA4} are satisfied, we have $\|\tbe_{nkm}\|=O_p\{(q_n\log\,n)^{1/2}\}$.	
\end{lemma}
\noindent \underline{Proof of Lemma \ref{lemma 0}}:
We calculate
\be
\|\tbe_{nkm}\|^2 &=& n\psi_{nkm}(\varepsilon)\trans X_{k\cdot a}(X_{k\cdot a}\trans B_{nkm}X_{k\cdot a})^{-2}X_{k\cdot a}\trans \psi_{nkm}(\varepsilon)  \nonumber \\
&\leq & \lambda_{\min} (n^{-1}X_{k\cdot a}\trans B_{nkm}X_{k\cdot a})^{-2}n^{-1}\psi_{nkm}(\varepsilon)\trans X_{k\cdot a}X_{k\cdot a}\trans \psi_{nkm}(\varepsilon)  \nonumber \\
&\leq & Cn^{-1}\psi_{nkm}(\varepsilon)\trans X_{k\cdot a}X_{k\cdot a}\trans \psi_{nkm}(\varepsilon)  \nonumber \\
&\leq & Cn^{-1}q_n (\hbox{$\max_{1\leq j\leq q_n}$}|\psi_{nkm}(\varepsilon)\trans X_{k\cdot j}|)^2 \nonumber \\
&=& Cn^{-1}q_n (\hbox{$\max_{1\leq j\leq q_n}$}|\hbox{$\si$}\psi_{kmi}(\varepsilon) X_{ki j}|)^2, \label{tbeta}
\ee	
where the third step uses Assumptions \ref{AA2} and \ref{AA3}. Since $\psi_{kmi}(\varepsilon)X_{kij}$ has mean zero and is bounded by Assumption \ref{AA1},
Hoeffding's inequality gives
\bse
\p\{|\hbox{$\si$}\psi_{kmi}(\varepsilon) X_{ki j}|\geq L_n(n\log\,n)^{1/2}\}\leq 2\exp\{-CL_n^2\log\,n\}
\ese
for any positive sequence $L_n\to\infty$. It follows that
\be
&& \phantom{=}\p\{\hbox{$\max_{1\leq j\leq q_n}$}|\hbox{$\si$}\psi_{kmi}(\varepsilon) X_{ki j}|\geq L_n(n\log\,n)^{1/2}\} \nonumber \\
&&\hskip 10mm \leq  \hbox{$\sum^{q_n}_{j=1}$}\p\{|\hbox{$\si$}\psi_{kmi}(\varepsilon) X_{ki j}|\geq L_n(n\log\,n)^{1/2}\} \nonumber \\
&&\hskip 10mm \leq  2q_n\exp\{-CL_n^2\log\,n\}=2q_n n^{-CL_n^2}\to 0,
\label{inprob}
\ee
where the last step holds true because $q_n=o(n^{1/2})$; see Assumption \ref{AA4}. Therefore
\bse
\hbox{$\max_{1\leq j\leq q_n}$}|\si \psi_{kmi}(\varepsilon) X_{ki j}|=O_p\{(n\log\,n)^{1/2}\}.
\ese
This combined with (\ref{tbeta}) gives $\|\tbe_{nkm}\|^2=O_p(q_n\log\,n)$, which completes the proof.
\\[3ex]
\noindent\underline{Proof of Theorem \ref{theorem 1}}:
	Under Assumptions \ref{AA1}-\ref{AA4}, Lemma 6 of \citet{sherwood2016partially} gives
\be
\|n^{1/2}(\hth_{km}-\theta^*_{km})-\tbe_{nkm}\|=o_p(1)
\label{s&w}
\ee
 for every $k$ and $m$, with $\tbe_{nkm}$ defined in Lemma \ref{lemma 0}. Therefore
\be
\|\hth_{km}-\theta^*_{km}\|=O_p\{n^{-1/2}(q_n \log\,n)^{1/2}\}.
\label{rate}
\ee
It follows that for every $k$ and $m$,
\bse
\hbox{$\max_{1\leq j\leq q_n}$}|\hat{\theta}_{kmj}-\theta^*_{kmj}|\leq\|\hat{\theta}_{k}-\theta^*_{k}\|=O_p\{n^{-1/2}(q_n \log\,n)^{1/2}\}=O_p\{n^{(c_1-1)/2}(\log\,n)^{1/2}\}.
\ese
Hence
\bse
	\max_{1\leq j\leq q_n}\|\hat{\theta}^{(j)}-\theta^{*(j)}\|_1\leq KM\max_{1\leq k\leq K}\max_{1\leq m\leq M}\max_{1\leq j\leq q_n}|\hat{\theta}_{kmj}-\theta^*_{kmj}|=O_p\{n^{(c_1-1)/2}(\log\,n)^{1/2}\},
\ese
which, combined with Assumption \ref{AA5}, yields
	\bse
	\hbox{$\min_{1\leq j\leq q_n}$}\|\hat{\theta}^{(j)}\|_1&\geq &\hbox{$\min_{1\leq j\leq q_n}$}\|\theta^{*(j)}\|_1-\hbox{$\max_{1\leq j\leq q_n}$}\|\hat{\theta}^{(j)}-\theta^{*(j)}\|_1 \\
	&\geq& C n^{(c_2-1)/2}-\{n^{(c_1-1)/2}(\log\,n)^{1/2}\}= O_p\{n^{(c_2-1)/2}\}.
	\ese
We assume $\lambda_n=o\{n^{(c_2-1)/2}\}$, which implies
\be
\p\{\hbox{$\min_{1\leq j\leq q_n}$}\|\hat{\theta}^{(j)}\|_1\geq a\lambda_n\}\to 1.
\label{thetaj}
\ee
The subderivative of the objective function (\ref{obhetero}) with respect to $\theta^{(j)}$ is 
\be
\frac{\partial \Gamma_{\lambda_n}(\theta)}{\partial \theta^{(j)}}=
\begin{cases}
	\partial \ell_n(\theta)/\partial \theta^{(j)}+\lambda_n\mathbb{S}(\theta^{(j)}),  &\|\theta^{(j)}\|_1\leq\lambda_n, \\
	\partial \ell_n(\theta)/\partial \theta^{(j)}+\mathbb{S}(\theta^{(j)})(a\lambda_n-\|\theta^{(j)}\|_1)/(a-1), &\lambda_n<\|\theta^{(j)}\|_1<a\lambda_n,  \\
	\partial \ell_n(\theta)/\partial \theta^{(j)}, & a\lambda_n\leq\|\theta^{(j)}\|_1, \label{diff}
\end{cases}
\ee
where $\mathbb{S}(\theta^{(j)})=(\hbox{Sign}(\theta_{11j}),\dots,\hbox{Sign}(\theta_{1Mj}),\dots,\hbox{Sign}(\theta_{K1j}),\dots,\hbox{Sign}(\theta_{KMj}))\trans$ with Sign$(x)=x/|x|$ for $x\neq 0$, and Sign$(0)=[-1,1]$. Thus (\ref{thetaj}) implies that, with probability tending to one, $\hth^{(j)}$ ($1\leq j\leq q_n$) belongs to the third case in (\ref{diff}). Combined with the fact that $\hth$ is a local minimizer of $\ell_n(\theta)$, it gives that
	\be
	0\in\partial \ell(\theta)/\partial \theta^{(j)} |_{\theta=\hth}=\partial \Gamma_{\lambda_n}(\theta)/\partial\theta^{(j)} |_{\theta=\hth}.
	\label{nonzero}
	\ee
	
	Under Assumptions \ref{AA1}-\ref{AA5}, Lemma 2.3 of \citet{wang2012quantile} yields that for every $k$ and $m$,
    \be
    \p\{\hbox{$\max_{q_n< j\leq p_n}$} |\partial \ell(\theta)/\partial\theta_{kmj} |_{\theta=\hth} |>\lambda_n\}\to 0.\label{zeropar}
	\ee
	Since $\|\hat{\theta}^{(j)}\|_1=0$ for $q_n<j\leq p_n$, which belongs to the first case in (\ref{diff}), we have
	\be
	\partial \Gamma_{\lambda_n}(\theta)/\partial\theta^{(j)} |_{\theta=\hth}=\partial \ell(\theta)/\partial \theta^{(j)} |_{\theta=\hth }+\lambda_n\mathbb{S}(\mathbf{0})  \label{firstcase}
	\ee
	Since $\mathbb{S}(\mathbf{0})=\{(u_1,\dots,u_{K}):|u_k|\leq 1, k=1\dots,K\}$, (\ref{zeropar}) and (\ref{firstcase}) imply that for $q_n<j\leq p_n$,
	\be
	\p\{0\in\partial \Gamma_{\lambda_n}(\theta)/\partial\theta^{(j)} |_{\theta=\hth}\}\to 1. \label{zero}
	\ee
		Combining (\ref{nonzero}) and (\ref{zero}) completes the proof.
%
\\[3ex]
\noindent\underline{Proof  of Theorem \ref{theorem 2}}:
Set $\hat{\beta}_n=n^{1/2}(\hat{\theta}_a-\theta_a^*)$, $\tilde{\beta}_n=n^{-1/2}R_n^{-1}X_a\trans\psi_n(\varepsilon)$ and write $A_n\Sigma_n^{-1/2}\tilde{\beta}_n=\si D_{ni}$, where $D_{ni}=n^{-1/2}A_n\Sigma^{-1/2}_nR_n^{-1}\delta_{ni}$, $\delta_{ni}=\{\psi_{1\cdot i}(\varepsilon)\trans\otimes X\trans_{1ia},\dots, \psi_{K\cdot i}(\varepsilon)\trans 
\otimes X\trans_{Kia}\}\trans$ and $\psi_{k\cdot i}(\varepsilon)=\{\psi_{k1i}(\varepsilon),\dots,\psi_{kMi}(\varepsilon)\}\trans$ for every $k$ and $i$. We have $E(D_{ni})=\bf{0}$ since $E(\delta_{ni})=\bf{0}$ and
	\bse
	\sumi E(D_{ni}D_{ni}\trans)&=&n^{-1}E[A_n\Sigma_n^{-1/2}R_n^{-1}\{\sumi E(\delta_{ni}\delta\trans_{ni}\mid\mathcal{X})\}R_n^{-1}\Sigma_n^{-1/2}A_n\trans] \\
	&=&E\{A_n\Sigma_n^{-1/2}R_n^{-1}(n^{-1}X_a\trans H_n X_a)R_n^{-1}\Sigma_n^{-1/2}A_n\trans\} \\
	&=&E(A_n\Sigma_n^{-1/2}R_n^{-1}S_n R_n^{-1}\Sigma_n^{-1/2}A_n\trans)=A_n A_n\trans\to G.
\ese
For any $\eta>0$ we obtain
	\bse
	\sumi E\{\|D_{ni}\|^2I(\|D_{ni}\|>\eta)\}&\leq & \eta^{-2}\sumi E(\|D_{ni}\|^4) \\
	&=&(n\eta)^{-2}\sumi E\{(\delta_{ni}\trans R_n^{-1}\Sigma_n^{-1/2}A_n\trans A_n\Sigma_n^{-1/2}R_n^{-1}\delta_{ni})^2\} \\	
          &\leq & (n\eta)^{-2}\lambda^2_{\max}(A_n\trans A_n)\sumi E\{(\delta_{ni}\trans R_n^{-1}\Sigma_n^{-1}R_n^{-1}\delta_{ni})^2\} \\
	&\leq& Cn^{-2}\sumi E\{(\delta_{ni}\trans S_n^{-1}\delta_{ni})^2\} \\
	&\leq & Cn^{-2}\sumi E\{\lambda_{\min}(S_n)^{-2}\|\delta_{ni}\|^4\} \\
	&\leq &Cn^{-2}\sumi E(\|\delta_{ni}\|^4) \\
	&=&Cn^{-2}\sumi E\{(\hbox{$\sk\sm$}\psi_{kmi}(\varepsilon)^2\|X_{kia}\|^2)^2\} \\
	&\leq & Cn^{-2}\sumi E\{(\hbox{$\max_{1\leq k\leq K}$}\|X_{kia}\|)^4\} \\
	&\leq & Cn^{-1}E\{(\hbox{$\max_{1\leq i\leq n}\max_{1\leq k\leq K}$}\|X_{kia}\|)^4\} \\
	&\leq&Cn^{-1}q_n^2=o(1),
          \ese
	with $\lambda_{\max}(\cdot)$ being the largest eigenvalue of a square matrix. The fourth step in the above display results from the fact that $\lambda_{\max}(A_n\trans A_n)\to C$. The sixth step uses the condition that $\lambda_{\min}(S_n)$ is uniformly bounded away from zero. The last but one step holds true because of Assumption \ref{AA1}, and the last step uses Assumption \ref{AA4}. This shows that the Lindeberg-Feller condition for the central limit theorem is satisfied, i.e.\ we have
	\be
	A_n\Sigma_n^{-1/2}\tilde{\beta}_n=\si D_{ni}\to N(0,G)\ \mbox{in distribution } (n \to \infty). \label{2.3}
	\ee
          It is obvious that $\tbe_{n}=(\tbe_{n11}\trans,\dots,\tbe_{n1M}\trans,\dots,\tbe_{nK1}\trans,\dots,\tbe_{nKM}\trans)\trans$ with $\tbe_{nkm}$ defined in Lemma \ref{lemma 0}. Hence, using (\ref{s&w}), we have
         \bse
\|\hbe_n-\tbe_n\|\leq \hbox{$\sk\sm$}\|\hat{\beta}_{nkm}-\tbe_{nkm}\|=o_p(1).
         \ese
It follows that
\bse
\|A_n\Sigma_n^{-1/2}(\hbe_n-\tbe_n)\|^2&=& (\hat{\beta}_n-\tbe_n)\trans\Sigma_n^{-1/2}A_n A_n\trans \Sigma_n^{-1/2}(\hat{\beta}_n-\tbe_n) \\
&\leq& \lambda_{\max}(A_n A_n\trans)\lambda_{\min}(\Sigma_n)^{-1}\|\hat{\beta}_n-\tbe_n\|^2=o_p(1).
\ese
In the last step we used $\lambda_{\max}(A_nA_n\trans)\to C$, Assumption \ref{AA2} and the condition that $\lambda_{\min}(S_n)$ is uniformly bounded away from zero. This combined with (\ref{2.3}) yields
\bse
n^{1/2}A_n\Sigma_n^{-1/2}(\hat{\theta}_a-\theta_a^*)=A_n\Sigma_n^{-1/2}\hat{\beta}_n\to N(0,G)\ \hbox{in distribution } (n\to\infty).
\ese

\begin{lemma}\label{lemma 2}
Set $\mm_1^*=\{\md:\md\in\mm,\md^*\subset\md\}$ and use the notation
     from Section \ref{sec3}. Let Assumptions \ref{AA1}, \ref{AA3}, \ref{AB2} and \ref{AB4} be satisfied. Let $c_4$ be the constant from Assumption \ref{AB4}. Then we have, for $k=1,\dots,K$, $m=1,\dots,M$, and any positive sequence $L_n$ satisfying $L_n\to\infty$ and $1\leq L_n(\log\,n)^{1/2}\leq n^{1/10-c_4/5}$,
	\bse
	\p\{|\si\{\rho_m(Y_{ki}-X_{ki\md}\trans\hth_{km\md})-\rho_m(\varepsilon_{kmi})\}|\leq L_n|\md|\log\,n,\ \hbox{for any $\md\in\mm^*_1$}\}\to 1.
	\ese
\end{lemma}
\noindent \underline{Proof of Lemma \ref{lemma 2}}: Under Assumptions \ref{AA1}, \ref{AA3}, \ref{AB2} and \ref{AB4}, Lemma A.2 in the supplement to \citet{lee2014model} gives
\be
\hbox{$\lim_{L\to\infty}\lim_{n\to\infty}$}\p\{\|\hth_{km\md}-\theta^*_{km\md}\|\leq Ln^{-1/2}(|\md|\log\,p_n)^{1/2},\ \hbox{for any $\md\in\mm_1^*$}\}= 1.
\label{distance}
\ee
Then, as $L_n\to\infty$,
\be
\p\{\|\hth_{km\md}-\theta^*_{km\md}\|\leq L_nn^{-1/2}(|\md|\log\,p_n)^{1/2},\ \hbox{for any $\md\in\mm_1^*$}\}\to 1.
\label{distance2}
\ee
Under Assumptions \ref{AA1}, \ref{AA3}, \ref{AB2} and \ref{AB4},
and since $1\leq L_n(\log\,n)^{1/2}\leq n^{1/10-c_4/5}$, we can apply Lemma A.1 in the supplement to \citet{lee2014model}, which gives
\be
\max_{\md\in\mm_1^*}\Big||\md|^{-1}[\hv-E(\hv\mid X_{k\cdot\md})+2\sum^n_{i=1}X_{ki\md}\trans(\hth_{km\md}-\theta^*_{km\md})\psi_{kmi}(\varepsilon)]\Big|=o_p(1)
\label{LeeA.1}
\ee
with $\hv=\si\{\rho_m(Y_{ki}-X_{ki\md}\trans\hth_{km\md})-\rho_m(\varepsilon_{kmi})\}$. Then we have, on an event that has probability tending to one,
\be
&&\phantom{=}|\si X_{ki\md}\trans(\hth_{km\md}-\theta^*_{km\md})\psi_{kmi}(\varepsilon)| \nonumber \\
&&\hskip 10mm \leq \|\hth_{km\md}-\theta^*_{km\md}\|\|\si X_{ki\md}\psi_{kmi}(\varepsilon)\| \nonumber \\
&&\hskip 10mm \leq \|\hth_{km\md}-\theta^*_{km\md}\||\md|^{1/2}\hbox{$\max_{1\leq j\leq p_n}$}|\si X_{kij}\psi_{kmi}(\varepsilon)| \nonumber \\
&&\hskip 10mm \leq L_n n^{-1/2}(|\md|\log\,p_n)^{1/2}|\md|^{1/2}L_n(n\log\,n)^{1/2}=L^2_n|\md|\log\,n
\label{xtheta}
\ee
for any $\md\in\mm_1^*$. The last but one step uses (\ref{inprob}) and (\ref{distance2}).
From Assumption \ref{AB4} we have $p_n=O(n^{c_3})$. Hence (\ref{inprob}) holds true when $q_n$ is substituted by $p_n$. We also have, for any $\theta_\md\in\mathbb{R}^{|\md|}$ satisfying $\|\theta_\md-\theta^*_{km\md}\|\leq L_nn^{-1/2}(|\md|\log\,p_n)^{1/2}$,
\be
&&\phantom{=}|\si E\{\rho_m(Y_{ki}-X_{ki\md}\trans\theta_{\md})-\rho_m(\varepsilon_{kmi})\mid X_{ki}\}| \nonumber \\
&&\hskip 10mm =\si E\{\hbox{$\int^{X\trans_{ki\md}(\theta_{\md}-\theta^*_{km\md})}_0$} I(\varepsilon_{kmi}\leq s)-I(\varepsilon_{kmi}\leq 0)ds\mid X_{ki}\} \nonumber \\
&&\hskip 10mm =\si \hbox{$\int^{X\trans_{ki\md}(\theta_{\md}-\theta^*_{km\md})}_0$} F_{km}(s\mid X_{ki})-F_{km}(0\mid X_{ki})ds \nonumber \\
&&\hskip 10mm =\si \hbox{$\int^{X\trans_{ki\md}(\theta_{\md}-\theta^*_{km\md})}_0$} sf_{km}(\bar{s}\mid X_{ki})ds \nonumber\\
&&\hskip 10mm \leq C(\theta_{\md}-\theta^*_{km\md})\trans\si (X_{ki\md}X_{ki\md}\trans)(\theta_{\md}-\theta^*_{km\md}) \nonumber\\
&&\hskip 10mm \leq Cn\lambda_{\max}(n^{-1}X_{k\cdot\md}\trans X_{k\cdot\md})\|\theta_\md-\theta^*_{km\md}\|^2 \nonumber\\
&&\hskip 10mm \leq Cn\|\theta_\md-\theta^*_{km\md}\|^2\leq CL_n^{2}|\md|\log\,p_n.
\label{integral}
\ee
The first step in the above results is from Knight's identity \citep{knight1998limiting}. In the second step, $F_{km}(\cdot\mid X_k)$ is the conditional distribution function of $\varepsilon_{km}$ given $X_k$. The third step uses a Taylor expansion with some $\bar{s}$ between $0$ and $X\trans_{ki\md}(\theta_{\md}-\theta^*_{km\md})$. The fourth step holds true because of Assumption \ref{AA3} and the fact that $\sup_{1\leq i\leq n}|X\trans_{ki\md}(\theta_{\md}-\theta^*_{km\md})|\leq \sup_{1\leq i\leq n}\|X_{ki\md}\|\|\theta_{\md}-\theta^*_{km\md}\|\leq CL_nd_nn^{-1/2}(\log\,n)^{1/2}\leq Cn^{4c_4/5-2/5}(\log\,n)^{1/2}\to 0$ (Assumptions \ref{AA1} and \ref{AB4}).  Combining (\ref{distance2}), (\ref{LeeA.1}), (\ref{xtheta}) and (\ref{integral}) yields that, for any $\md\in\mm^*_1$,
\bse
\hv&\leq&|E(\hv\mid X_{k\cdot\md})|+2|\si X_{ki\md}\trans(\hth_{km\md}-\theta^*_{km\md})\psi_{kmi}(\varepsilon)|+|\md|o_p(1) \\
&\leq&CL_n^{2}|\md|\log\,p_n+L^2_n|\md|\log\,n+|\md|o_p(1)
\leq CL_n^2|\md|\log\,n
\ese
 with probability approaching one, where the $o_p(1)$ term comes from (\ref{LeeA.1}). This finishes the proof.
\\[3ex]
\noindent\underline{Proof of Theorem \ref{theorem 3}}:
Consider the set of overfitted models $\mm_1=\{\md\in\mathcal{M}:\md^*\subset\md,\md\neq\md^*\}$ and the set of underfitted models $\mathcal{M}_2=\{\md\in\mathcal{M}:\md^*\not\subset\md\}$. Since $\mm_1\cup\mm_2=\mm\backslash\{\md^*\}$ it suffices to show
	\be
	&&\hbox{$\lim_{n\to\infty}$}\p\{\hbox{$\min_{\md\in\mathcal{M}_1}$}\hbox{MQBIC}(\md)>\hbox{MQBIC}(\md^*)\}= 1, \label{over} \\
	&&\hbox{$\lim_{n\to\infty}$}\p\{\hbox{$\min_{\md\in\mathcal{M}_2}$}\hbox{MQBIC}(\md)>\hbox{MQBIC}(\md^*)\}= 1. \label{under}
	\ee
	
	We first prove (\ref{over}). Write $\hw_{\md}=\nn\sk\sm\si\rho_m(Y_{ki}-X_{ki\md}\trans\hth_{km\md})$ and $W^*=\nn\sk\sm\si\rho_m(\varepsilon_{kmi})$. From Lemma \ref{lemma 2} we know that we can choose some sequence $L_n$ that does not depend on $\md$ and satisfies $L_n\to\infty$, $L_n=o(T_n)$ and $n^{-1}L_nd_n\log\,n\to 0$ such that for $k=1,\dots,K$ and $m=1,\dots,M$,
	\be
	&&\p\{|\si\{\rho_m(Y_i-X_{ki\md}\trans\hth_{km\md})-\rho_m(\varepsilon_{kmi})\}| \nonumber \\
	&&\phantom{\p\{} \leq (MK)^{-1}L_n |\md|\log\,n,\ \hbox{for any $\md\in\mm^*_1$}\}\to 1.
	\label{A13}
	\ee
	Since $|\hw_{\md}-W^*|\leq \nn\hbox{$\sk\sm$}|\si\{\rho_m(Y_i-X_{ki\md}\trans\hth_{km\md})-\rho_m(Y_i-X_{ki\md^*}\trans\theta^*_{km\md^*})\}|$ we have $\p\{|\hw_{\md}-W^*|\leq \nn L_n|\md|\log\,n,\ \hbox{for any $\md\in\mm_1^*$}\}\to 1$. It follows that
	\be
	\p\{|\hw_\md-\hw_{\md^*}|\leq n^{-1}L_n(|\md|+|\md^*|)\log\,n,\ \hbox{for any $\md\in\mm^*_1$}\}\to 1
	\label{rhodiff2}
	\ee
	and that, for some positive constants $C_5$ and $C_6$,
	\be
	\p\{C_5\leq\hw_{\md^*}\leq C_6,\  \hbox{for any $\md\in\mm_1^*$}\}\to 1.
	\label{denominator}
	\ee
Here we used Assumption \ref{AB6} and the fact that $n^{-1}L_n|\md^*|\log\,n\to 0$ (Assumption \ref{AB4}). Therefore, with probability tending to one,
	\be
	&&\phantom{=}\hbox{$\min_{\md\in\mathcal{M}_1}$}\hbox{MQBIC}(\md)-\hbox{MQBIC}(\md^*) \nonumber \\
&&=\hbox{$\min_{\md\in\mathcal{M}_1}$}[\log\{1+\hw_{\md^*}^{-1}(\hw_\md-\hw_{\md^*})\}+(2n)^{-1}T_n(|\md|-|\mathcal{D^*}|)\log\,n ] \nonumber \\
&&\geq\hbox{$\min_{\md\in\mathcal{M}_1}$}\{-2\hw_{\md^*}^{-1}|\hw_\md-\hw_{\md^*}|+(2n)^{-1}T_n(|\md|-|\mathcal{D^*}|)\log\,n\} \nonumber \\
&&\geq\hbox{$\min_{\md\in\mathcal{M}_1}$}\{-Cn^{-1}L_n(|\md|+|\mathcal{D^*}|)\log\,n+(2n)^{-1}T_n(|\md|-|\mathcal{D^*}|)\log\,n\}. \label{A15}
	\ee
The first inequality in the above derivation comes from the fact that $\log(1+x)\geq -2|x|$ for any $|x|\in(-1/2,1/2)$, from (\ref{rhodiff2}) combined with $n^{-1}L_nd_n\log\,n\to 0$, and from (\ref{denominator}). The last step holds true because of (\ref{rhodiff2}) and (\ref{denominator}). Then  (\ref{A15}) implies (\ref{over}) because $L_n=o(T_n)$ and $|\md|>|\md^*|$.
	
	To prove equation (\ref{under}) we introduce $\md'=\md\cup\md^*$ for any $\md\in\mathcal{M}_2$. Since $q$ is fixed by Assumption \ref{AB4}, there is a parameter with minimum absolute value $\nu >0$, i.e.\ $\nu=\min_{1\leq k\leq K}\min_{1\leq m\leq M}\min_{j\in\md^*}|\theta_{kmj}^*|>0$. Since (\ref{distance}) still holds for any set in $\mm_2^*=\{\md\subset\{1, \dots, p_n\}:|\md|\leq 2d_n,\md^*\subset\md\}$, we have
	\be
	\p\{\hbox{$\max_{\md\in\mm_2}$}\|\hth_{km\md'}-\theta^*_{km\md'}\|\leq\nu\}\to 1.
	\label{leq}
	\ee
	For $k=1,\dots,K$, $m=1,\dots,M$ and any $\md\in\mm_2$, let $\tth$ be a $|\D'|\times 1$ vector, i.e.\ the dimension of $\tth$ is given by the number of indices in the set $\D' = \D \cup \D^*$.
	We define it as an extended version of $\hth_{km\md}$:
	the components of $\tth$ that correspond to the index set $\D$ coincide
	with the components of $\hth_{km\md}$; the remaining components are filled with zeros.
	For example, if $\D = \{1,3\}$, $\D^* = \{1,2\}$ and $\hth_{km\md} = \{1.4, 0.7\}$,
	then $\D' = \{1,2,3\}$, $|\D'|=3$ and $\tth = (1.4, 0, 0.7)\trans$. Since $\md^*\not\subset\md$, there exist some $k_0$ and $m_0$ such that $\|\ttho-\theta^*_{k_0m_0\md'}\|\geq\nu$. Combined with (\ref{leq}) and since the check function is convex, this implies that there exists a $|\md'|\times 1$ vector $\bth$ such that $\|\bth-\theta^*_{k_0m_0\md'}\|=\nu$ and $$\si\rho_{m_0}(Y_{k_0i}-X_{k_0i\md'}\trans\bth)\leq\si\rho_{m_0}(Y_{k_0i}-X_{k_0i\md'}\trans\ttho)=\si\rho_{m_0}(Y_{k_0i}-X_{k_0i\md}\trans\hth_{k_0m_0\md}).$$
	Write $B_\nu(\md')=\{\omega\in\mathbb{R}^{|\md'|}:\|\omega\|=\nu\}$ and $G_{\md'}(\omega)=\nn\sum^n_{i=1}\{\rho_{m_0}(\varepsilon_{k_0m_0i}-X_{k_0i\md'}\trans\omega)-\rho_{m_0}(\varepsilon_{k_0m_0i})\}$. Then we have, for any $\md\in\mathcal{M}_2$,
    \be
	&&\phantom{=} \nn\si\{\rho_{m_0}(Y_{k_0i}-X_{k_0i\md}\trans\hth_{k_0m_0\md})-\rho_{m_0}(Y_{k_0i}-X_{k_0i\md'}\trans\hth_{k_0m_0\md'})\} \nonumber \\
	&&\geq \nn\si\{\rho_{m_0}(Y_{k_0i}-X_{k_0i\md}\trans\bth)-\rho_{m_0}(Y_{k_0i}-X_{k_0i\md'}\trans\hth_{k_0m_0\md'})\} \nonumber \\
	&&=G_{\md'}(\bth-\theta^*_{k_0m_0\md'})-G_{\md'}(\hth_{k_0m_0\md'}-\theta^*_{k_0m_0\md'})+ \nonumber\\
	&& \phantom{=}E\{G_{\md'}(\bth-\theta^*_{k_0m_0\md'})\mid X_{k_0\cdot\md'}\}-E\{G_{\md'}(\bth-\theta^*_{k_0m_0\md'})\mid X_{k_0\cdot\md'}\} \nonumber\\
	&&\geq\hbox{$\inf_{\omega\in B_{\nu}(\md')}$}E\{G_{\md'}(\omega)\mid X_{k_0\cdot\md}\}-\hbox{$\sup_{\omega\in B_{\nu}(\md')}$} |G_{\md'}(\omega)-E\{G_{\md'}(\omega)|X_{k_0\cdot\md'}\}|- \nonumber \\
	&&\phantom{=}G_{\md'}(\hth_{k_0m_0\md'}-\theta^*_{k_0m_0\md'}). \label{A20}
    \ee
	Similar to the calculation of (\ref{integral}) we have, for any $\md'\in\mm^*_2$ and $\omega\in B_\nu(\md')$,
	\be
	E\{G_{\md'}(\omega)\mid X_{k_0\cdot\md'}\}&=&\nn\si \hbox{$\int^{X_{k_0i\md'}\trans\omega}_0$} F_{k_0m_0}(s\mid X_{k_0i\md'})-F_{k_0m_0}(0\mid X_{k_0i\md'})ds \nonumber \\
	&=&\nn\si \hbox{$\int^{X\trans_{k_0i\md'}\omega}_0$} sf_{k_0m_0}(\bar{s}\mid X_{k_0i\md'})ds \nonumber\\
	&\geq& C\omega\trans\{\nn\si (X_{k_0i\md'}X_{k_0i\md'}\trans)\}\omega \nonumber\\
	&\geq& C\lambda_{\min}(n^{-1}X_{k_0\cdot\md'}\trans X_{k_0\cdot\md'})\|\omega\|^2=C\|\omega\|^2,
	\label{A21}
	\ee
	where the third step uses Assumption (\ref{AA3}) and the last step Assumption (\ref{AB2}). Then, under Assumptions \ref{AA1}, \ref{AA3}, \ref{AB2} and \ref{AB4}, Lemma A.3 in the supplement to \citet{lee2014model} gives
	\be
	\hbox{$\max_{\md'\in\mathcal{M}_2^*}\sup_{\omega\in B_{\nu}(\md')}$} |G_{\md'}(\omega)-E\{G_{\md'}(\omega)\mid X_{k_0\cdot\md'}\}|=o_p(1). \label{A22}
	\ee
	It is obvious that (\ref{A13}) is still valid when $\mm_1^*$ is substituted by $\mm_2^*$. Hence
	\bse
	\pr\{\hbox{$\max_{\md'\in\mm_2^*}$} |G_{\md'}(\hth_{k_0m_0\md'}-\theta^*_{k_0m_0\md'})|\leq Cn^{-1}L_nd_n\log\,n\}\to 1,
	\ese
	which gives $\max_{\md'\in\mm_2^*}|G_{\md'}(\hth_{k_0m_0\md'}-\theta^*_{k_0m_0\md'})|=o_p(1)$. This, combined with (\ref{A20}), (\ref{A21}) and (\ref{A22}) implies that, with probability approaching one,
	\be
	\nn\hbox{$\min_{\md\in\mm_2}$}\si\{\rho_m(Y_{k_0i}-X_{k_0i\md}\trans\hth_{k_0m_0\md})-\rho_m(Y_{k_0i}-X_{k_0i\md'}\hth_{k_0m_0\md'})\}\geq 2C.
	\label{rhodiff3}
	\ee
    Since $\md\in\md'$ we have $\si\{\rho_m(Y_{ki}-X_{ki\md}\trans\hth_{km\md})-\rho_m(Y_{ki}-X_{ki\md'}\hth_{km\md'})\}\geq 0$ for any $k$, $m$ and $\md\in\mm_2$. It follows
	\bse
	\hw_\md-\hw_{\md'}&=&\nn\sk\sm\si\{\rho_m(Y_{ki}-X_{ki\md}\trans\hth_{km\md})-\rho_m(Y_{ki}-X_{ki\md'}\hth_{km\md'})\} \\
	&\geq& \nn\si\{\rho_m(Y_{k_0i}-X_{k_0i\md}\trans\hth_{k_0m_0\md})-\rho_m(Y_{k_0i}-X_{k_0i\md'}\hth_{k_0m_0\md'})\}.
	\ese
	This, combined with (\ref{rhodiff3}), gives
	\be
	\p\{\hbox{$\min_{\md\in\mm_2}$}(\hw_\md-\hw_{\md'})\geq 2C\}\to 1.
	\label{wdiff}
	\ee
	 Then, with probability tending to one,
	\be
		&&\phantom{=} \hbox{$\min_{\md\in \mathcal{M}_2}$}\hbox{MQBIC}(\md)-\hbox{MQBIC}(\md') \nonumber \\
		&&\hskip 10mm =\hbox{$\min_{\md\in \mathcal{M}_2}$} [\log\{1+\hw_{\md'}^{-1}(\hw_\md-\hw_{\md'})\}-(2n)^{-1}T_n(|\md'|-|\md|)\log\,n ] \nonumber \\
		&&\hskip 10mm \geq \hbox{$\min_{\md\in \mathcal{M}_2}$} [\min \{\log\, 2,\hw_{\md'}^{-1}(\hw_\md-\hw_{\md'})/2\}-(2n)^{-1}T_n|\md^*|\log\,n ] \nonumber \\
		&&\hskip 10mm \geq\hbox{$\min_{\md\in \mathcal{M}_2}$} [\min \{\log\, 2,\hw_{\md'}^{-1}C\}-(2n)^{-1}T_n|\md^*|\log\,n ]>0 \label{A17}
	\ee
The first inequality comes from the fact that $\log(1+x)\geq\min\{x/2,\log\, 2\}$ for any $x\geq 0$. The second inequality uses (\ref{wdiff}). The last step uses Assumption \ref{AB5} and the fact that (\ref{denominator}) is still valid when $\mm_1^*$ is substituted by $\mm_2^*$. Since (\ref{over}) can be easily extended to any $\md\in(\mathcal{M}_2^*\backslash\{\md^*\})$, we know that, with probability tending to one, MQBIC$(\md')\geq$MQBIC$(\md^*)$ for any $\md'\in\mm_2^*$. This and (\ref{A17}) yield
	\bse
	&&\phantom{=}\hbox{$\min_{\md\in \mathcal{M}_2}$}\hbox{MQBIC}(\md)-\hbox{MQBIC}(\md^*)\\
	&&\hskip 10mm =\hbox{$\min_{\md\in \mathcal{M}_2}$}\{\hbox{MQBIC}(\md)-\hbox{MQBIC}(\md')+\hbox{MQBIC}(\md')-\hbox{MQBIC}(\md^*)\} \\
	&&\hskip 10mm \geq \hbox{$\min_{\md\in \mathcal{M}_2}$}\{\hbox{MQBIC}(\md)-\hbox{MQBIC}(\md')\}>0,
	\ese
with probability tending to one. This proves (\ref{under}).

\bibliographystyle{biometrika}
\bibliography{myreference-di}

\end{document}